\documentclass[onecolumn,secnumarabic,amssymb, amsmath, nofootinbib,tightenlines,
nobibnotes, aps, prl,epsfig]{revtex4}
\usepackage{graphicx}
\usepackage{dcolumn}
\usepackage{bm}
\begin{document}
\preprint{APS/123-QED}
\title{The dipole cross section by the unintegrated gluon distribution at small $x$ }

\author{G.R.Boroun}%
 \email{boroun@razi.ac.ir }
\affiliation{ Department of Physics, Razi University, Kermanshah
67149, Iran}
\date{\today}
\begin{abstract}
We apply a previously developed scheme to consistently include the
improved saturation model for the unintegrated gluon distribution
(UGD) in order to evaluate, in the framework of $k_{t}$
factorization, at small $x$ at the next-to-leading order (NLO) in
$\alpha_{s}$. We start the unintegrated gluon distribution with a
parametrization of the deep inelastic structure function for
electromagnetic scattering with protons, and then extract the
color dipole cross section, which preserves its behavior success
in a wide range of $k_{t}^{2}$ in comparisons with the UGD models
(M. Hentschinski, A. Sabio Vera and C. Salas (HSS), I.P.Ivanov and
N.N.Nikolaev (IN)  and G. Watt, A.D. Martin and M.G. Ryskin (WMR))
. These results show that the geometric scaling holds for the
improved saturation model in a wide kinematic region $rQ_{s}$, and
 are comparable with the Golec-Biernat-W$\ddot{\mathrm{u}}$sthoff (GBW) model.
 The unintegrated gluon distribution at low and high momentum transfer in a wide
 range of $x$ is considered.\\

\end{abstract}
 \pacs{***}
\maketitle
\subsection{I. Introduction}

Although our knowledge of the proton structure at small-$x$ is
very limited, novel opportunities will be opened at new-generation
facilities (Electron-Ion Collider(EIC), High-Luminosity Large
Hadron Collider (HL-LHC), Forward Physics Facility (FPF)).
Combining the information coming from dipole cross sections and
$p_{T}$-unintegrated densities could play an important role. In
particular, polarized amplitudes and cross sections for the
exclusive electroproduction of $\rho$ and $\phi$ mesons at the
Hadron-Electron Ring Accelerator (HERA) and the EIC are very
sensitive to the unintegrated gluon distribution (UGD) model
adopted, whereas forward Drell-Yan dilepton distributions at the
Large Hadron Collider beauty (LHCb) are very sensitive to
next-to-leading logarithmic corrections. Indeed, the dipole cross
section is directly connected via a Fourier transform to the
small-x UGD, whose evolution in $x$ is regulated by the Balitsky-
Fadin-Kuraev-Lipatov (BFKL) equation [1]. The BFKL equation
retains the full $Q^2$ dependence and not just the leading
$\ln{Q^2}$ terms. Indeed the resummation of terms is proportional
to $\alpha_{s}\ln(1/x)$ to all orders. This involves considering
this means that we do not have strongly ordered $k_{t}$ but
instead integrate over the full range of $k_{t}$ in phase space of
the gluons [2,3]. The BFKL equation governs the evolution of the
UGD,
 where the $k_{t}$-factorization is used in the high energy limit in which the QCD
interaction is described in terms of the quantity which depends on
the transverse momentum of the gluon [4]. In the
$k_{t}$-factorization framework, the gluon distribution depends on
$x$ and $k_{t}^{2}$, where $x$ and $k^{2}_{t}$ being the
fractional momentum of proton carried by gluon and the transverse
momentum of gluon respectively. At high energies, the
$k_{t}$-factorization is a suitable formalism to compute the
relevant distribution and cross sections. Within this regime, the
longitudinal momentum fraction of partons, $x$, is small. The
unintegrated gluon distribution function is thus of great
phenomenological and theoretical interest to develop the formalism
which includes the transverse momentum dependence in the context
of the multigluon distributions.\\
It is more appropriate to use the parton distributions
unintegrated over the transverse momentum $k_{t}$ in the framework
of $k_{t}$-factorization QCD approach for semi-inclusive processes
(such as inclusive jet production in DIS, electroweak boson
production, etc.) at high energies. The $k_{t}$-factorization
formalism provides solid theoretical grounds for the effects of
initial gluon radiation and intrinsic parton transverse momentum
$k_{t}$. The unintegrated  gluon $f(x, k_{t}^{2})$ distribution is
directly related to the dipole-nucleon cross section, that is
saturated at low $Q$ or large transverse distances $r{\sim}1/Q$
between quark $q$ and antiquark $\overline{q}$ in the
$q\overline{q}$ dipole created from the splitting of the virtual
photon $\gamma^{*}$ in the ep DIS. Indeed, the suitable
factorization approach in DIS (where
$W{\gg}Q{\gg}\Lambda_{\mathrm{QCD}}$) is provided by
$k_{t}$-factorization [5].\\
The UGD, in its original definition, obeys the BFKL  evolution
equation in the $x$ variable and being a nonperturbative quantity.
To realistically describe the structure of the proton, we must
introduce a $k_{t}$ unintegrated gluon density, whose evolution at
small-$x$ is governed by the BFKL equation [6]. The object of the
BFKL evolution equation at very small $x$ is the differential
gluon structure function\footnote{Eq.(1) is modified with the
Sudakov form factor as $x$ increase.} of proton
\begin{eqnarray}
f(x,k_{t}^{2})&=&\frac{\partial{xg(x,\mu^{2})}}{\partial{\ln}\mu^{2}}|_{\mu^{2}=k^{2}_{t}}
\end{eqnarray}
which emerges in the color dipole picture (CDP) of inclusive deep
inelastic scattering (DIS)  and diffractive DIS into dijets [7].
Unintegrated distributions are required to describe measurements
where transverse momenta are exposed explicitly.\\
The unintegrated gluon distribution function satisfies the BFKL
equation for an alternative derivation in terms of color dipoles.
The BFKL equation at leading order is given by
\begin{eqnarray}
\frac{{\partial}f(x,k^{2}_{t})}{{\partial}{\ln}({1}/{x})}=\int{dk'^{2}_{t}}K(k^{2}_{t},k'^{2}_{t})f(x,k'^{2}_{t}),
\end{eqnarray}
which describes the evolution in ${\ln}({1}/{x})$ of the
unintegrated gluon density and $K$ is the BFKL kernel  [2,3]. In
the small $x$ limit this basically gives a power law behavior in
$x$,
\begin{eqnarray}
f(x,k^{2}_{t}){\sim}h(k^{2}_{t})x^{-\lambda},
\end{eqnarray}
where $h{\sim}(k^{2}_{t})^{-\frac{1}{2}}$ at large $k^{2}_{t}$ and
$\lambda$ is the maximum eigenvalue of the kernel $K$ of the BFKL
equation. For fixed $\alpha_{s}$, $\lambda$ has the value
$\lambda=\frac{\alpha_{s}}{\pi}12\ln2$, where this hard Pomeron
has been termed the BFKL Pomeron and lead to very steeply rising
$V^{*}N$ cross-sections. In the BFKL analysis, there are infra-red
(IR) and ultra-violet (UV) cutoffs on the $k^{2}_{t}$
integrations. Indeed determining the IR cutoff parameter is
important for the integrating
${\int}dk^{2}_{t}\alpha_{s}(k^{2}_{t})$ down to $k^{2}_{t}=0$,
also the choice of the UV cutoff is important when working at
finite order\footnote{This is the reason for applying that the
DGLAP formulation ensures energy conservation order by order, but
the BFKL formulation does not [2,3].} [8,9].\\
The color dipole picture (CDP) [10] has been introduced to study a
wide variety of small $x$ inclusive and diffractive processes at
HERA. The CDP, at small $x$, gives a clear interpretation of the
high-energy interactions, where is characterized by high gluon
densities because the proton structure is dominated by dense gluon
systems [11-13] and predicts that the small $x$ gluons in a hadron
wavefunction should form a Color Glass Condensate [14]. Dipole
representation provides a convenient description of DIS at small
$x$. There, the scattering between the virtual photon $\gamma^{*}$
and the proton is seen as the
 color dipole where the transverse dipole size $r$ and the
 longitudinal momentum fraction $z$ with respect to the photon
 momentum are defined. The amplitude for the complete process is simply the production of
these subprocess amplitudes, as the DIS cross section is
factorized into a light-cone wave function and a dipole cross
section. Using the optical theorem, this leads to the following
expression for the $\gamma^{*}p$ cross-sections
\begin{eqnarray}
\sigma_{L,T}^{\gamma^{*}p}(x,Q^{2})=\int dz d^{2}\mathbf{r}
|\Psi_{L,T}(\mathbf{r},z,Q^{2})|^{2}\sigma_{\mathrm{dip}}({x},\mathbf{r}),
\end{eqnarray}
where subscripts $L$ and  $T$ referring to the transverse and
longitudinal polarization state of the exchanged boson. Here
$\Psi_{L,T}$ are the appropriate spin averaged light-cone wave
functions of the photon and $\sigma_{\mathrm{dip}}({x},r)$ is the
dipole cross-section which related to the imaginary part of the
$(q\overline{q})p$ forward scattering amplitude. The variable $z$,
with $0\leq z \leq 1 $, characterizes the distribution of the
momenta between quark and antiquark. The square of the photon wave
function describes the probability for the occurrence of a
$(q\overline{q})$ fluctuation of transverse size with respect to
the photon polarization.\\
Another framework which can be used for calculating the parton
distributions is based on the
Dokshitzer-Gribov-Lipatov-Altarelli-Parisi (DGLAP) evolution
equations [15]. Deep inelastic electron-proton scattering is
described in terms of scale dependent parton densities
$q(x,\mu^{2})$ and $g(x,\mu^{2})$ [16], where the integrated gluon
distribution ($xg(x,\mu^{2})$) is defined through the unintegrated
gluon distribution ($f(x,k_{t}^{2})$) by
\begin{eqnarray}
xg(x,\mu^{2})&{\equiv}&\int^{\mu^{2}}\frac{dk_{t}^{2}}{k_{t}^{2}}f(x,k_{t}^{2}).
\end{eqnarray}
The unintegrated gluon distribution is related to the dipole cross
section [4, 17]
\begin{eqnarray}
\sigma(x,\mathbf{r})=\frac{8\pi^{2}}{N_{c}}\int\frac{dk_{t}}{k_{t}^{^{3}}}[1-J_{0}(k_{t}\mathbf{r})]\alpha_{s}f(x,k_{t}^{2}).
\end{eqnarray}
A novel formulation of the UGD for DIS in a way that accounts for
the leading powers in both the Regge and Bjorken limits is
presented in Ref.[18]. In this way, the UGD is defined by an
explicit dependence on the longitudinal momentum fraction $x$
which entirely spans both the dipole operator and the gluonic
Parton Distribution Function (PDF).\\
In addition to the gluon momentum derivative model (i.e., Eq.(1)),
several other models [7,10, 19-21] for the UGD have also been
proposed so far. A comparison between these models can be found in
Refs.[22,23]. The authors in Ref. [19] presented an
$x$-independent model (ABIPSW) of the UGD where merely coincides
with the proton impact factor by the following form
\begin{eqnarray}
f(x,k_{t}^{2})=\frac{A}{4\pi^{2}M^{2}}\Big{[}\frac{k_{t}^{2}}{M^{2}+k_{t}^{2}}\Big{]},
\end{eqnarray}
where M is a characteristic soft scale and A is the normalisation
factor.\\
The authors in Ref. [7] presented a UGD soft-hard model (IN) in
the large and small $k_{t}$ regions by the following form
\begin{eqnarray}
f(x,k_{t}^{2})=f_{\mathrm{soft}}^{(B)}(x,k_{t}^{2})\frac{k_{s}^{2}}{k_{s}^{2}+k_{t}^{2}}
+f_{\mathrm{\mathrm{hard}}}(x,k_{t}^{2})\frac{k_{t}^{2}}{k_{h}^{2}+k_{t}^{2}},
\end{eqnarray}
where the soft and the hard components are defined in [7].\\
The UGD model was considered in [20] to used in the study of DIS
structure functions and takes the form of a convolution between
the BFKL gluon Green$^{,}$s function and a leading-order (LO)
proton impact factor, where has been employed in the description
of single-bottom quark production at LHC and to investigate the
photoproduction of $J/\Psi$ and $\Upsilon$, by the following form
(HSS model)
\begin{eqnarray}
f(x,k_{t}^{2},M_{{h}})&=&\int_{-\infty}^{+\infty}\frac{d\nu}{2\pi^{2}}\mathcal{C}
\frac{\Gamma(\delta-i\nu-\frac{1}{2})}{\Gamma(\delta)}(\frac{1}{x})^{\chi(\frac{1}{2}+i\nu)}
(\frac{k_{t}^{2}}{Q_{0}^{2}})^{\frac{1}{2}+i\nu}\Bigg{\{}1+\frac{\overline{\alpha}_{s}^{2}\beta_{0}\chi_{0}(\frac{1}{2}+i\nu)}{8N_{c}}\log(\frac{1}{x})\nonumber\\
&&{\times}\Bigg{[}-\psi(\frac{1}{2}+i\nu)-\log\frac{k_{t}^{2}}{M^{2}_{{h}}}\Bigg{\}}\Bigg{]},
\end{eqnarray}
where $\chi_{0}(\frac{1}{2}+i\nu)$ and $\chi(\gamma)$ are
respectively the LO and the next-to-leading order (NLO)
eigenvalues of the BFKL kernel and $\beta_{0}=11-\frac{2}{3}n_{f}$
with $n_{f}$ the number of active quarks. The LO eigenvalue of the
BFKL kernel is
$\chi_{0}(\frac{1}{2}+i\nu){\equiv}\chi_{0}(\gamma)=2\psi(1)-\psi(\gamma)-\psi(1-\gamma)$
and the NLO eigenvalue of the BFKL kernel is
$\chi(\gamma)=\overline{\alpha_{s}}\chi_{0}(\gamma)+\overline{\alpha_{s}}^{2}\chi_{1}(\gamma)
-\frac{1}{2}\overline{\alpha_{s}}^{2}\chi'_{0}(\gamma)\chi_{0}(\gamma)
+\chi_{RG}(\overline{\alpha_{s}},\gamma)$ with $\psi(\gamma)$ is
the logarithmic derivative of the Euler Gamma function. Here
$\overline{\alpha}_{s}=\frac{3}{\pi}\alpha_{s}(\mu^{2})$ with
$\mu^{2}=Q_{0}M_{h}$ where $M_{h}$ plays the role of the hard
scale which can be identified with the photon virtuality,
$\sqrt{Q^{2}}$.\\
The authors in Ref. [21] presented a UGD model (Watt-Martin-Ryskin
(WMR) model) where depends on an extra-scale $\mu$, fixed at $Q$,
by the following form
\begin{eqnarray}
f(x,k_{t}^{2},\mu^{2})&=&T_{g}(k_{t}^{2},\mu^{2})\frac{\alpha_{s}(k_{t}^{2})}{2\pi}
\int_{x}^{1}dz\Bigg{[}\sum_{q}P_{gq}(z)\frac{x}{z}q(\frac{x}{z},k_{t}^{2})
+P_{gg}(z)\frac{x}{z}g(\frac{x}{z},k_{t}^{2})\Theta(\frac{\mu}{\mu+k_{t}}-z)
\Bigg{]},
\end{eqnarray}
where $T_{g}(k_{t}^{2},\mu^{2})$ gives the probability of evolving
from the scale $k_{t}$ to the scale $\mu$ without parton emission
and $P_{ij}^{,}$s are the splitting functions.\\
Golec-Biernat-Wusthoff (GBW) [10] presented a UGD model where
derives from the effective dipole cross section
$\sigma(x,\mathbf{r})$ for the scattering of a $q\overline{q}$
pair of a nucleon as\footnote{The reader can be referee to
Refs.[7,10, 19-21] for a meticulous treatment of the parameters.}
\begin{eqnarray}
f(x,k_{t}^{2})&=&k_{t}^{4}\sigma_{0}\frac{R_{0}^{2}(x)}{2\pi}e^{-k_{t}^{2}R_{0}^{2}(x)},
\end{eqnarray}
with
$R_{0}^{2}(x)=\frac{1}{\mathrm{GeV}^{2}}(\frac{x}{x_{0}})^{\lambda_{p}}$
and the following values $\sigma_{0}=23.03~\mathrm{mb}$,
$\lambda_{p}=0.288$ and $x_{0}=3.04{\times}10^{-4}$. Although one
of them (the HSS one) was fitted to reproduce DIS structure
functions, the study of other reactions has provided an evidence
that the UGD is not yet well known. The HSS model also reproduces
well the forward Drell-Yan data at the LHC without any further
adjustment of extra parameters [24]. In this paper we use the
DGLAP-improved saturation model with respect to the UGD in the
proton to access the dipole cross
section at low $x$.\\

\subsection{II. Method}

The interaction of the $q\overline{q}$ pair with the proton is
described by the dipole cross section. It is related to the gluon
density in the target by the $k_{T}$-factorization formula
[7,10,17]
\begin{eqnarray}
\sigma(x,r)=\int\frac{d^{2}k_{t}}{k_{t}^{^{4}}}\frac{\partial{xg(x,k^{2}_{t})}}{\partial{\ln}k^{2}_{t}}
[1-e^{i\mathbf{k}_{t}.\mathbf{r}}][1-e^{-i\mathbf{k}_{t}.\mathbf{r}}],
\end{eqnarray}
where the relation between $f(x,k^{2}_{t})$ and $xg(x,k^{2}_{t})$
defined through Eq.(1). In the following we present a method of
extraction of the gluon distribution function in the kinematic
region of low values of the Bjorken variable $x$ from the
structure $F_{2}(x,k^{2}_{t})$ and derivative
$\partial{F_{2}(x,k^{2}_{t})}/\partial{\ln}k^{2}_{t}$ by relying
on the DGLAP $Q^{2}$-evolution equations. The structure function
$F_{2}$ is expressed via the singlet and gluon distributions as
\begin{eqnarray}
F_{2}(x,k^{2}_{t})=<e^2>\bigg{[}B_{2,s}(x){\otimes}xf_{s}(x,k^{2}_{t})+B_{2,g}(x){\otimes}xg(x,k^{2}_{t})\bigg{]},
\end{eqnarray}
where $<e^{2}>$ is the average of the charge $e^{2}$ for the
active quark flavors $n_{f}$ ,
$<e^{2}>=n_{f}^{-1}\sum_{i=1}^{n_{f}}e_{i}^{2}$ and the nonsinglet
densities become negligibly small in comparison with the singlet
densities at small $x$. The quantities $B_{2,i}(x) (i=s,g)$  are
the known Wilson coefficient functions and the symbol $\otimes$
denotes convolution according to the usual prescription. According
to the DGLAP evolution equations, the
 singlet distribution function leads to the following relation of integro-differential
 equation [25]
\begin{eqnarray}
\frac{{\partial}F_{2}(x,k^{2}_{t})}{{\partial}{\ln}k^{2}_{t}}&=&-\frac{a_{s}(k^{2}_{t})}{2}[P_{qq}(x){\otimes}F_{2}(x,k^{2}_{t})
+<e^{2}>P_{qg}(x){\otimes}xg(x,k^{2}_{t})],
\end{eqnarray}
where
\begin{eqnarray}
P_{a,b}(x)=P_{a,b}^{(0)}(x)
+a_{s}(k^{2}_{t})\widetilde{P}_{a,b}^{(1)}(x)+a_{s}^{2}(k^{2}_{t})\widetilde{P}_{a,b}^{(2)}(x)
\end{eqnarray}
and
\begin{eqnarray}
\widetilde{P}_{ab}^{(n)}(x)={P}_{ab}^{(n)}(x)+[C_{2,s}+C_{2,g}+...]\otimes
{P}_{ab}^{(0)}(x)+... .\nonumber
\end{eqnarray}
The quantities $\widetilde{P}_{ab}$$^{,}s$ are expressed via the
known splitting  and Wilson coefficient functions  in literatures
[26,27] and
$a_{s}(k^{2}_{t})=\alpha_{s}(k^{2}_{t})/4\pi$.\\
Considering the variable definitions $\upsilon{\equiv}\ln(1/x)$
and $w{\equiv}\ln(1/z)$, one can rewrite Eq.(14) in terms of the
convolution integrals and new variables\footnote{For further
discussions please see Ref.[28]. } as
\begin{eqnarray}
\frac{\partial{\mathcal{\widehat{F}}_{2}(\upsilon,k^{2}_{t})}}{\partial{\ln}k^{2}_{t}}&=&\int_{0}^{\upsilon}[\mathcal{\widehat{F}}_{2}(\upsilon,k^{2}_{t})
\mathcal{\widehat{H}}^{(\varphi)}_{2,s}(a_{s}(k^{2}_{t}),\upsilon-w)
+<e^{2}>\mathcal{\widehat{G}}(\upsilon,k^{2}_{t})
\mathcal{\widehat{H}}^{(\varphi)}_{2,g}(a_{s}(k^{2}_{t}),\upsilon-w)]dw,
\end{eqnarray}
where
\begin{eqnarray}
\frac{\partial{\mathcal{\widehat{F}}_{2}(\upsilon,k^{2}_{t})}}{\partial{\ln}k^{2}_{t}}{\equiv}
\frac{{\partial}F_{2}(e^{-\upsilon},k^{2}_{t})}{\partial{\ln}k^{2}_{t}},~
\mathcal{\widehat{G}}(\upsilon,k^{2}_{t}){\equiv}xg(e^{-\upsilon},k^{2}_{t}),~
\mathcal{\widehat{H}}^{(\varphi)}(a_{s}(k^{2}_{t}),\upsilon){\equiv}e^{-\upsilon}\widehat{P}_{a,b}^{(\varphi)}(a_{s}(k^{2}_{t}),\upsilon).
\end{eqnarray}
The splitting function reads
\begin{eqnarray}
 P_{a,b} ^{(\varphi)}
(a_{s},x)=\sum_{n=0}^{\varphi}a_{s}^{n+1}(k^{2}_{t})P_{a,b}^{(n)}(x),
\end{eqnarray}
where $n$ denotes the order in running coupling
$\alpha_{s}(k^{2}_{t})$. The Laplace transform of
$\mathcal{\widehat{H}}(a_{s}(k^{2}_{t}),\upsilon)$$^{,}s$
 are given by the following
 forms
\begin{eqnarray}
\Phi_{f}^{(\varphi)}(a_{s}(k^{2}_{t}),s)&{\equiv}&
{\mathcal{L}}[\mathcal{\widehat{H}}^{(\varphi)}_{2,s}(a_{s}(k^{2}_{t}),\upsilon);s]
=\int_{0}^{\infty}\mathcal{\widehat{H}}^{(\varphi)}_{2,s}(a_{s}(k^{2}_{t}),\upsilon)e^{-s\upsilon}d\upsilon,\nonumber\\
 \Theta_{f}^{(\varphi)}(a_{s}(k^{2}_{t}),s)&{\equiv}&{\mathcal{L}}[\mathcal{\widehat{H}}^{(\varphi)}_{2,g}(a_{s}(k^{2}_{t}),\upsilon);s]
 =\int_{0}^{\infty}\mathcal{\widehat{H}}^{(\varphi)}_{2,g}(a_{s}(k^{2}_{t}),\upsilon)e^{-s\upsilon}d\upsilon.
\end{eqnarray}
We know that the Laplace transforms of the convolution factors are
simply the ordinary products of the Laplace transforms of the
factors. Therefore, Eq.(16) in the Laplace space $s$ reads as
\begin{eqnarray}
\frac{\partial{f_{2}(s,k^{2}_{t})}}{\partial{\ln}k^{2}_{t}}&=&
\Phi_{f}^{(\varphi)}(a_{s}(k^{2}_{t}),s)f_{2}(s,k^{2}_{t})+<e^{2}>\Theta_{f}^{(\varphi)}(a_{s}(k^{2}_{t}),s)xg(s,k^{2}_{t}),
\end{eqnarray}
where
\begin{eqnarray}
{\mathcal{L}}[\mathcal{\widehat{F}}_{2}(\upsilon,k^{2}_{t});s]=f_{2}(s,k^{2}_{t}),~
{\mathcal{L}}[\mathcal{\widehat{G}}(\upsilon,k^{2}_{t});s]=xg(s,k^{2}_{t}).
\end{eqnarray}
The gluon distribution into the parametrization of the proton
structure function and its derivative with respect to
${\ln}k^{2}_{t}$ in $s$-space in Eq.(20) is given by the following
form
\begin{eqnarray}
xg(s,k^{2}_{t})&=&k^{(\varphi)}(a_{s}(k^{2}_{t}),s)Df_{2}(s,k^{2}_{t})-h^{(\varphi)}(a_{s}(k^{2}_{t}),s)f_{2}(s,k^{2}_{t}),
\end{eqnarray}
where
\begin{eqnarray}
Df_{2}(s,k^{2}_{t})&=&{\partial{f_{2}(s,k^{2}_{t})}}/{\partial{\ln}k^{2}_{t}},\nonumber\\
k^{(\varphi)}(a_{s}(k^{2}_{t}),s)&=&1/(<e^{2}>\Theta^{(\varphi)}_{f}(a_{s}(k^{2}_{t}),s)),\nonumber\\
h^{(\varphi)}(a_{s}(k^{2}_{t}),s)&=&\Phi^{(\varphi)}_{f}(a_{s}(k^{2}_{t}),s)k^{(\varphi)}(a_{s}(k^{2}_{t}),s).\nonumber
\end{eqnarray}
The inverse Laplace transforms of Eq.(22) reads
\begin{eqnarray}
\widehat{xg}^{(\varphi)}(\upsilon,k^{2}_{t}){\equiv}{\mathcal{L}}^{-1}[xg(s,k^{2}_{t});\upsilon]
={\mathcal{L}}^{-1}[k^{(\varphi)}(a_{s}(k^{2}_{t}),s)Df_{2}(s,k^{2}_{t})-h^{(\varphi)}(a_{s}(k^{2}_{t}),s)f_{2}(s,k^{2}_{t});\upsilon]
\end{eqnarray}
where the inverse transform of a product to the convolution of the
original functions, giving
\begin{eqnarray}
{\mathcal{L}}^{-1}[f(s){\times}h(s);\upsilon]&=&\int_{0}^{\upsilon}\widehat{F}
(w)\widehat{H}(\upsilon-w)dw.\nonumber
\end{eqnarray}
The inverse Laplace transform of the functions $k$ and $h$ in
Eq.(23) are defined by ${\widehat{J}}(\upsilon)$ and
${\widehat{M}}(\upsilon)$, as
\begin{eqnarray}
\widehat{J}(\upsilon,k^{2}_{t}){\equiv}{\mathcal{L}}^{-1}[k(s,k^{2}_{t});\upsilon],
~
\widehat{M}(\upsilon,k^{2}_{t}){\equiv}{\mathcal{L}}^{-1}[h(s,k^{2}_{t});\upsilon].
\end{eqnarray}
The functions $k$ and $h$ in $s$-space, into the Laplace transform
of the splitting functions ($P_{qq}$ and $P_{qg}$), are given by
[29]
\begin{eqnarray}
k^{(n)}(a_{s}(k^{2}_{t}),s)&=&\bigg{(}<e^{2}>{\mathcal{L}}[e^{-\upsilon}\sum_{n=0}^{1}a_{s}^{n+1}(k^{2}_{t})\widehat{P}_{qg}^{(n)}(\upsilon);s]\bigg{)}^{-1},\nonumber\\
h^{(n)}(a_{s}(k^{2}_{t},s)&=&\frac{{\mathcal{L}}[e^{-\upsilon}\sum_{n=0}^{1}a_{s}^{n+1}(k^{2}_{t})\widehat{P}_{qq}^{(n)}(\upsilon);s]}
{<e^{2}>{\mathcal{L}}[e^{-\upsilon}\sum_{n=0}^{1}a_{s}^{n+1}(k^{2}_{t})\widehat{P}_{qg}^{(n)}(\upsilon);s]},
\end{eqnarray}
where at small $x$, the kernels at NLO approximation in Eq.(19)
are given by the following forms
\begin{eqnarray}
\Theta_{f}(a_{s},s)&{\simeq}&2n_{f}a_{s}{\Big{[}}\frac{1}{1+s}-\frac{2}{2+s}+\frac{2}{3+s}{\Big{]}}
+a^{2}_{s}C_{A}T_{f}{\Big{[}}\frac{40}{9s}{\Big{]}},\nonumber\\
\Phi_{f}(a_{s},s)&{\simeq}&a_{s}{\Big{[}}4-\frac{8}{3}(\frac{1}{1+s}+\frac{1}{2+s}+2S_{1}(s){\Big{]}}
+a^{2}_{s}C_{F}T_{f}{\Big{[}}\frac{40}{9s}{\Big{]}},
\end{eqnarray}
where $S_{1}(s)=\psi(s+1)+\gamma_{E}$, where $\psi(x)$ is the
digamma function and $\gamma_{E}=0.5772156 . . .$ is Euler
constant. Here $C_{A}=N_{c}=3$,
$C_{F}=\frac{N_{c}^{2}-1}{2N_{c}}=\frac{4}{3}$ and
 $T_{f}=\frac{1}{2}n_{f}$. The standard representation for QCD
couplings in NLO (within the $\mathrm{\overline{MS}}$-scheme)
approximation reads
\begin{eqnarray}
\alpha^{\mathrm{NLO}}_{s}(k^{2}_{t})&=&\frac{4\pi}{\beta_{0}\ln\frac{k^{2}_{t}}{\Lambda^{2}}}\Big{[}1
-\frac{\beta_{1}}{\beta_{0}^{2}}\frac{\ln{\ln\frac{k^{2}_{t}}{\Lambda^{2}}}}{\ln\frac{k^{2}_{t}}{\Lambda^{2}}}\Big{]},
\end{eqnarray}
where $\beta_{0}$ and $\beta_{1}$ are the one and two loop
correction to the QCD $\beta$-function and $\Lambda$ is the QCD
cut-off parameter\footnote{The running coupling (27) is used in
the evolution of the DGLAP equations. In the HSS approach, a
running consistent with a global fit to jet observables was used
to find the correct form for the DIS structure functions at small
values of $x$ in the full $Q$ range [22]. In this regard, the
infrared freezing of strong coupling at leading order (LO) is
imposed by fixing $\Lambda_{QCD}=200~\mathrm{MeV}$ as
$\alpha_{s}(\mu^2)=\mathrm{min}\bigg{\{}0.82,\frac{4\pi}{\beta_{0}
\ln(\frac{\mu^2}{\Lambda^{2}_{QCD}})}\bigg{\}} $}.\\
Therefore, the integrated gluon density is related to the proton
structure function by the following form
\begin{eqnarray}
xg^{(\varphi)}(x,k^{2}_{t})&=&\int_{x}^{1}\Big{[}{DF}_{2}
(\ln\frac{1}{y},k^{2}_{t}){J}(a_{s}(k^{2}_{t}),\ln\frac{y}{x})
-{F}_{2}
(\ln\frac{1}{y},k^{2}_{t}){M}(a_{s}(k^{2}_{t}),\ln\frac{y}{x})\Big{]}\frac{dy}{y},
\end{eqnarray}
where $F_{2}(x,k^{2}_{t})$ is the proton structure function. The
parametrization of $F_{2}$ in Ref.[30] has an expression for the
asymptotic part of $F_{2}$ (no-valence) that accounts for the
asymptotic behavior $F_{2}{\simeq}\ln^{2}(1/x)$ at small $x$ which
describes fairly well the available experimental data on the
reduced
cross sections.\\
In next section we consider the UGD and the color dipole cross
section due to the parametrization of the proton structure
function in Eqs.(1) and (12) with respect to (28), respectively.\\


\subsection{3. Numerical Results}

The UGD is obtained directly in terms of the parameterization of
the structure function $F_{2}(x,k^{2}_{t})$ and its derivative.
The resulting UGD with the $k^{2}_{t}$-dependence due to the
parametrization of the proton structure function are shown in Fig.
1. In this figure, we plot the $k^{2}_{t}$ dependence of the UGD
at $x=10^{-3}$ and compared the unintegrated gluon distribution
behavior due to the improved saturation model with the GBW model.
An enhancement and then depletion is observable in the improved
and GBW models. These picks occur at
$k^{2}_{t}{\approx}10~\mathrm{GeV}^{2}$ and
$k^{2}_{t}{\approx}1~\mathrm{GeV}^{2}$ in the improved and GBW
models respectively. As we can seen, the UGD behavior in the
improved saturation model, in a wide range of $k^{2}_{t}$, is
softer than the GBW model. The error bands in Fig.1 in the
improved saturation model are due to the uncertainties in the
coefficients of the parametrization of the proton structure
function. The uncertainties are very small for low $k_{t}^{2}$ and
increases as $k_{t}^{2}$ increase to $100~\mathrm{GeV}^2$.\\
\begin{figure}[h]
\centerline{
\includegraphics[width=0.6\textwidth]{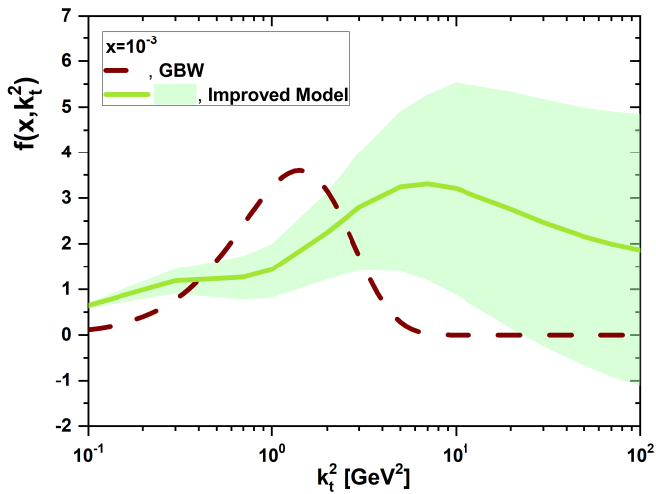}}
\caption{The UGD as a function of $k_{t}^{2}$ with respect to the
improved model compared with the GBW model at $x=10^{-3}$. The
error bands in the improved model are due to the statistical
uncertainties in the coefficients of the parametrization $F_{2}$
in [30].}\label{Fig1}
\end{figure}
\begin{figure}[h]
\centerline{
\includegraphics[width=0.6\textwidth]{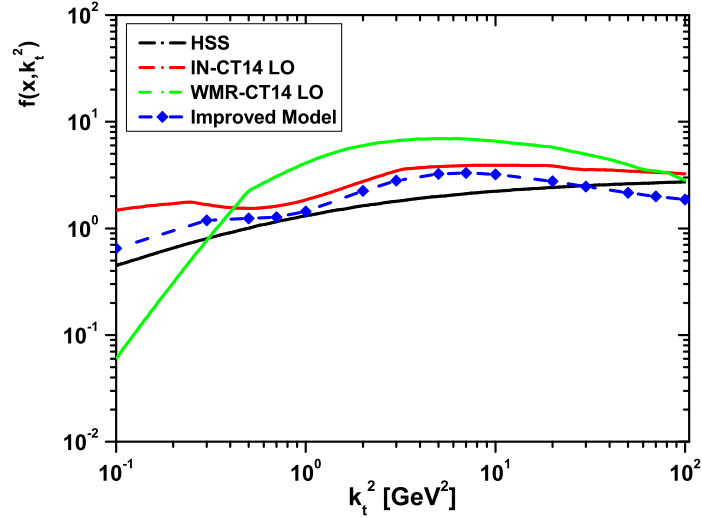}}
\caption{The unintegrated gluon distribution $f(x,k^{2}_{t})$
obtained from the improved saturation model (dashed-symbol), as a
function of $k^{2}_{t}$ at $x=10^{-3}$, compared with the HSS
[20], IN [7] and WMR [21] models.}\label{Fig2}
\end{figure}
\begin{figure}[h]
\centerline{
\includegraphics[width=0.6\textwidth]{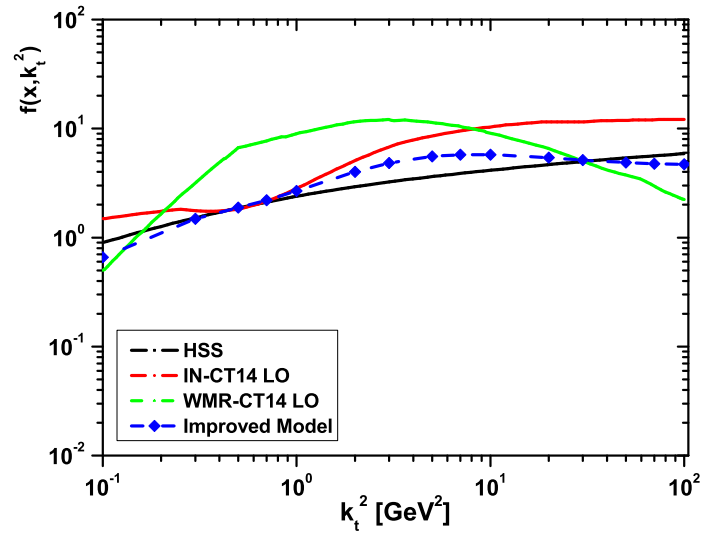}}
\caption{The same as Fig.2 for $x=10^{-4}$.}\label{Fig3}
\end{figure}
In Fig. 2 we show the $k_{t}$ distributions of three different
unintegrated gluons at $x=10^{-3}$. This figure compares the
results of the improved saturation model, based on the
parametrization of the proton structure function, with the HSS
[20], IN [7] and WMR [21] models. We observe that the improved
saturation result is comparable with the HSS and IN models in a
wide range of $k^{2}_{t}$. The differences are not large, however
there is some suppression due to the models at large and small
values of $k^{2}_{t}$. The continuous behavior of the UGD in our
model with increases of $k^{2}_{t}$ is due to the gluon and quark
terms included in the improved saturation model.\\
\begin{figure}[h]
\centerline{
\includegraphics[width=0.6\textwidth]{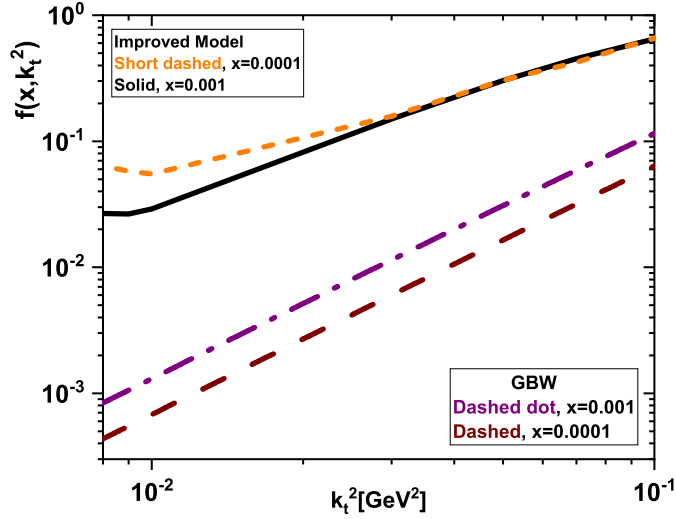}}
\caption{Low $k^{2}_{t}$-behavior of the considered UGD model for
$x=10^{-3}$ and $10^{-4}$ (solid and short dashed curves) compared
with the GBW model (dashed-dot and dashed curves) [10],
respectively.}\label{Fig4}
\end{figure}
\begin{figure}[h]
\centerline{
\includegraphics[width=0.6\textwidth]{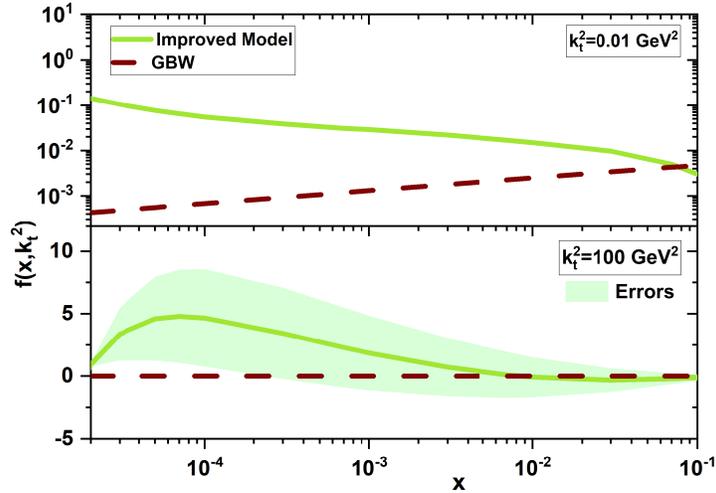}}
\caption{$x$-behavior of the considered UGD model (solid curves)
for $k^{2}_{t}=0.01~\mathrm{GeV}^2$ (up) and
$k^{2}_{t}=100~\mathrm{GeV}^2$ (down), as accompanied with the
statistical uncertainties from the coefficients of the
parameterization of $F_{2}$ in [30], compared with the GBW model
(dashed curves) [10].}\label{Fig5}
\end{figure}
 In Fig.3 we present the unintegrated gluons with the $k_{t}^{2}$-dependence of all the considered UGD
models in Fig.2, for $x=10^{-4}$. The plot clearly shows the same
behavior in the $k_{t}^{2}$-shape of the figures at low
values of $x$. In conclusion, the consistency of the several UGDs in their $k_{t}$ dependence
 holds for values of $x=10^{-3}$ and $10^{-4}$ in a wide range of
 $k_{t}$.
Calculation of the dipole cross section requires the knowledge of
the unintegrated gluon density for all scales
$0<k_{t}^{2}<\infty$. Usually the unintegrated gluon density is
known for $k_{t}^{2}>k_{t0}^{2}$ ($k_{t0}^{2}=1~\mathrm{GeV}^2$),
so it is interesting to consider the function $f(x,k_{t}^{2})$ for
lowest values of $k_{t}^{2}<k_{t0}^{2}$. Figure (4) illustrates
the unintegrated gluon distributions $f(x,k_{t}^{2})$ at low
$k_{t}^{2}$
($10^{-2}~\mathrm{GeV}^2{\lesssim}k_{t}^{2}{\leq}10^{-1}~\mathrm{GeV}^2$).
It shows that the modified UGD is different from the original GBW
UGD at $k_{t}^{2}{\lesssim}10^{-2}~\mathrm{GeV}^2$ and it similar
to the GBW UGD at $k_{t}^{2}>10^{-2}~\mathrm{GeV}^2$ with a higher
rate. In this range, the difference between the GBW UGD, for two
different values of the longitudinal momentum fraction,
$x=10^{-3}$ and $10^{-4}$ is uniform, while they are coincide with
the improved UGD for
$3{\times}10^{-2}~\mathrm{GeV}^2{\lesssim}k_{t}^{2}{\leq}10^{-1}~\mathrm{GeV}^2$.
These models in Fig.4 fairly reflect the distinct approaches
whence each UGD descends in the range
$3{\times}10^{-2}~\mathrm{GeV}^2{\lesssim}k_{t}^{2}{\leq}10^{-1}~\mathrm{GeV}^2$.
In Fig.5 we investigate the effect of different values of
$k_{t}^{2}$ on the unintegrated gluon distribution in a wide range
of $x$ and compared with the GBW UGD model. We observe that the
result for low $k_{t}^{2}$ (i.e., $k_{t}^{2}=0.01~\mathrm{GeV}^2$)
is different with the GBW UGD model in the range $x<0.1$. The
difference between the results increases as $x$ decreases. In
particular the sensitivity of the predictions to a detailed
parametrization of the infrared region which satisfies the gauge
invariance constraints as $k_{t}^{2}{\rightarrow}0$ is considered.
In Fig.5 (down figure) we compare our results with the GBW UGD
model at $k_{t}^{2}=100~\mathrm{GeV}^2$. The behavior of the GBW
UGD model is uniform in a wide range of $x$ and it is almost zero
in this range. Our results has a fluctuation in the region
$10^{-5}<x<10^{-3}$ where this is the largest discrepancy is
observed. This is due to the fact that as the gluon momentum
fraction $x$ decreases, the probability of the gluon splitting
increases. One can see that the improved UGD is different from the
GBW UGD at $10^{-5}<x<10^{-3}$ and coincides with it at larger $x$
($10^{-3}<x<10^{-1}$). Indeed, we have shown that our UGD is
similar to the GBW UGD, obtained at large $k_{t}$ (within the
uncertainties) and different from it at low $k_{t}$ in a wide
range of $x$. The error bands in the improved model are due to the
statistical uncertainties in the coefficients of the
parametrization $F_{2}$
in [30].\\
 In Fig.6, we have
calculated the improved saturation model with respect to the
unintegrated gluon distribution to the ratio
$\sigma_{\mathrm{dip}}/\sigma_{0}$ in a wide range of the dipole
size at the NLO approximation. Results of calculations and
comparison with the GBW model [10]  for $x=10^{-3}$ and $10^{-4}$
are presented in figures 6 and 7, respectively. The effective
parameters in the GBW model have been extracted from a fit of the
HERA data as $\lambda=0.288$ and $x_{0}=3.04{\times}10^{-4}$.
These corrections to the ratio of color dipole cross sections at
NLO approximation are comparable with the GBW model. For
$rQ_{s}{\geq}1$,  the dashed curve (central values) merge due to
geometric scaling in the dipole cross section in this region. In
the right-hand of Figs.6 and 7, a particular interests present the
ratio $\sigma_{\mathrm{dip}}/\sigma_{0}$ defined by the scaling
variable $rQ_{s}$, which means that the scattering amplitude and
corresponding cross sections can scale on the dimensionless scale
$rQ_{s}$. In conclusion, we observe that the geometric scaling
holds for other values of $x$ in a wide range of $rQ_{s}$. We
observe that its violation for $rQ_{s}<1$ is visible as $x$ decreases.\\
\begin{figure}[h]
\centerline{
\includegraphics[width=0.6\textwidth]{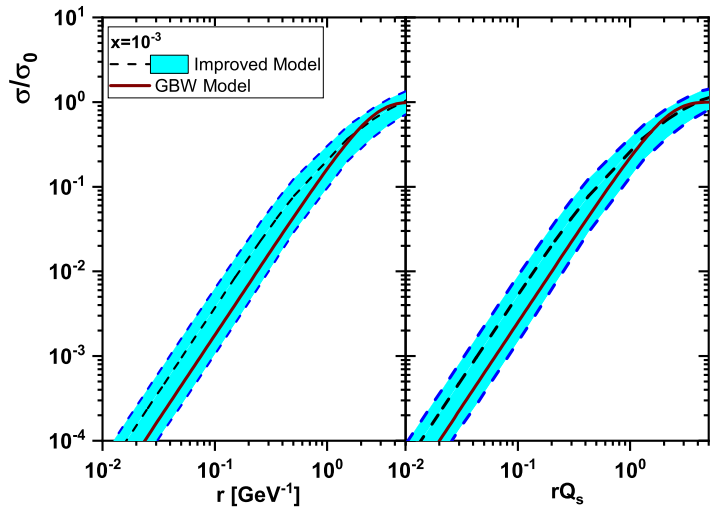}}
\caption{The ratio of the dipole cross sections in the improved
saturation model (dashed lines) as a function of $r$ (left plot)
and $rQ_{s}$ (right plot) for $x=10^{-3}$ at the NLO
approximation. The solid lines are due to the GBW model [10]. The
error bands are due to the statistical uncertainties from the
coefficients of the parameterization of $F_{2}$ in
[30].}\label{Fig6}
\end{figure}
\begin{figure}[h]
\centerline{
\includegraphics[width=0.6\textwidth]{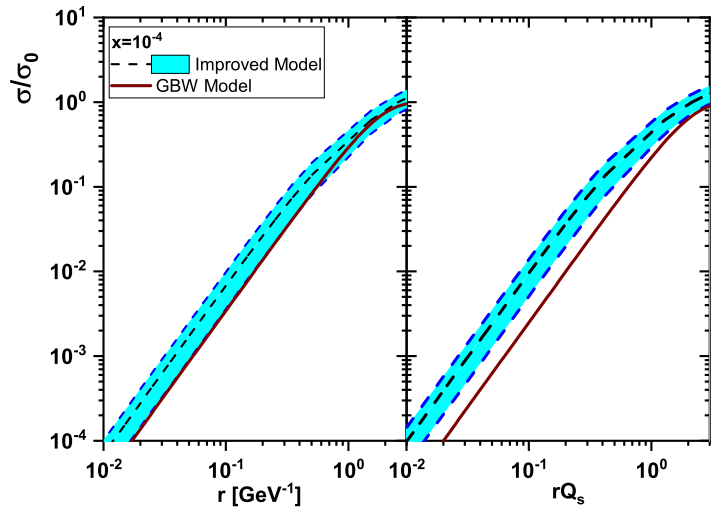}}
\caption{The same as Fig.6 for $x=10^{-4}$.}\label{Fig7}
\end{figure}
In summary, we study the unintegrated gluon distribution from a
parameterization of the proton structure function. In this
analysis we present the $k_{t}^{2}$-dependence of the improved UGD
model at the NLO approximation and compare with four models for
$f(x,k_{t}^{2})$, which exhibit rather different shape of
$k_{t}$-dependence in the region,
$10^{-1}{\leq}k_{t}^{2}{\leq}10^{2}~ \mathrm{GeV}^{2}$, for a
value of the longitudinal momentum fraction, $x=10^{-3}$. We show
the behaviour of our predictions for the UGD are comparable with
HSS and IN models in a wide range of $k_{t}^{2}$. These results
are comparable with the WMR and GBW models at moderate values of
$k_{t}^{2}$. It was shown that the improved UGD is different from
the  GBW UGD at low $k_{t}$ and it coincides with the GBW UGD at
large $k_{t}$.\\
Then we have presented the improved dipole cross section when the
unintegrated gluon distribution is derived from the
parametrization of the proton structure function as a function of
$r$ and $rQ_{s}$, respectively. The results according to the UGD
are consistent with the GBW saturation model at the NLO
approximation. The error bars are due to the statistical
uncertainties of the effective parameters and preserved that the
NLO results give a reasonable data description in comparison with
the GBW model. In this method, the large dipole size part of the
dipole cross section retains the features of the GBW model with
the saturation scale. The dipole cross section at the small dipole
size by the presence of the uninetrgated gluon distribution is
modified in comparison with the saturation scale of the GBW saturation model.\\

\subsection{ACKNOWLEDGMENTS}
The author is grateful to Razi University for the financial
 support of this project. I am also very grateful to the respectable reviewer
  of the $Eur.Phys.J.C {\bf82}, 740 (2022)$ article for suggesting this topic. Thanks also go to Z.Asadi for help
with preparation of the UGD model.\\

%

\section{References}
1. V.S.Fadin, E.A.Kuraev and L.N.Lipatov, Phys.Lett.B \textbf{60},
50(1975); L.N.Lipatov, Sov.J.Nucl.Phys. \textbf{23}, 338(1976);
I.I.Balitsky and L.N.Lipatov, Sov.J.Nucl.Phys.
\textbf{28}, 822(1978).\\
2. G.Dissertori, I.G.Knowles and M.Schmelling, Quantum
Chromodynamics High Energy Experiments and Theory, Oxford
University Press, 2009; R.K.Ellis, W.J.Stirling and B.R.Webber,
QCD and Collider Physics, Cambridge University Press, 1996.\\
3. A.M.Cooper-Sarkar, R.C.E.Devenish and A. De Roeck,
Int.J.Mod.Phys.A{\bf13}, 3385 (1998).\\
4. K.Kutak and A.M.Stasto, Eur.Phys.J.C {\bf41}, 343 (2005).\\
5. G.I.Lykasov, A.A.Grinyuk and V.A.Bednyakov, arXiv
[hep-ph]:1301.5156.\\
6. A.D.Bolognino, F.G.Celiberto, M.Fucilla, Dmitry Yu. Ivanov,
A.Papa, W.Schafer  and A.Szczurek, arXiv[hep-ph]:2202.02513.\\
7. I.P.Ivanov and N.N.Nikolaev, Phys.Rev.D {\bf65}, 054004
(2002).\\
8. J.Kwiecinski, A.D.Martin and P.J.Sutton, Z.Phys.C {\bf71}, 585
(1996); Jeffrey R.Forshaw, P.N.Harriman and P.J.Sutton,
J.Phys.G {\bf19}, 1616 (1993).\\
9. A.J.Askew, J.Kwiecinski, A.D.Martin and P.J.Sutton, Phys.Rev.
D{\bf49}, 4402 (1994); E.Elias,
K.Golec-Biernat and  Anna M.Stasto, JHEP {\bf01}, 141 (2018).\\
10. K.Golec-Biernat and  M.W$\ddot{\mathrm{u}}$sthoff, Phys. Rev.
D {\bf59},  014017 (1998); K. Golec-Biernat and S.Sapeta, JHEP
{\bf03}, 102 (2018).\\
11. J.Bartels, K.Golec-Biernat and H.Kowalski, Phys. Rev.
D{\bf66},
014001 (2002).\\
12. B.Sambasivam, T.Toll and T.Ullrich, Phys.Lett.B {\bf803}, 135277 (2020).\\
13. J.R.Forshaw and G.Shaw, JHEP {\bf12},
052 (2004).\\
14. E.Iancu, A.Leonidov and L.McLerran, Nucl.Phys.A {\bf692}, 583
(2001); Phys.Lett.B {\bf510}, 133 (2001); E.Iancu,K.Itakura and
S.Munier, Phys.Lett.B {\bf590}, 199
(2004).\\
15. Yu.L.Dokshitzer, Sov.Phys.JETP{\bf46}, 641 (1977); G.Altarelli
and G.Parisi, Nucl.Phys.B {\bf126}, 298 (1977); V.N.Gribov and
L.N.Lipatov, Sov.J.Nucl.Phys. {\bf15},
438 (1972).\\
16. M.A.Kimber, J.Kwiecinski, A.D.Martin and A.M.Stasto,
Phys.Rev.D
{\bf62}, 094006 (2000).\\
17. N.N.Nikolaev and B.G.Zakharov, Phys.Lett.B {\bf332}, 184
(1994);
 N. N. Nikolaev and W. Sch$\ddot{a}$fer, Phys. Rev. D {\bf74}, 014023 (2006).\\
18. R.Boussarie and Y.Mehtar-Tani, Phys.Lett.B {\bf831}, 137125
(2022).\\
19. I.V. Anikin, A. Besse, D.Yu. Ivanov, B. Pire, L. Szymanowski
and S. Wallon, Phys. Rev. D {\bf84}, 054004 (2011).\\
20. M. Hentschinski, A. Sabio Vera and C. Salas, Phys. Rev. Lett.
{\bf110},  041601 (2013).\\
21. G. Watt, A.D. Martin and M.G. Ryskin, Eur. Phys. J. C {\bf31},
73
(2003).\\
22. A.D.Bolognino, F.G.Celiberto, Dmitry Yu. Ivanov and A.Papa,
arXiv [hep-ph]:1808.02958; arXiv [hep-ph]:1902.04520; arXiv
[hep-ph]:1808.02395.\\
23. F.G.Celiberto, Nuovo Cim. C {\bf42}, 220 (2019).\\
24. F.G.Celiberto, D. Gordo Gomez and A.Sabio Vera, Phys.Lett.B
{\bf786}, 201 (2018).\\
25. L.P.Kaptari, A.V.Kotikov, N.Yu.Chernikova, and P.Zhang,
Phys.Rev.D {\bf99}, 096019 (2019).\\
26. J. Blumlein, V. Ravindran and W. van Neerven, Nucl. Phys. B
\textbf{586}, 349(2000); S.Catani and F.Hautmann,
Nucl.Phys.B{\bf427}, 475(1994).\\
27. D.I.Kazakov and A.V.Kotikov, Phys.Lett.B{\bf291}, 171(1992);
E.B.Zijlstra and W.L.van Neerven, Nucl.Phys.B{383}, 525(1992).\\
28. M.M. Block, L.Durand, P.Ha, D.W.McKay, Eur.Phys.J.C {\bf 69},
425 (2010).\\
29. G.R.Boroun, Eur.Phys.J.C {\bf82}, 740 (2022); G.R.Boroun, Eur.Phys.J.C {\bf83}, 42 (2023).\\
30. M. M. Block, L. Durand and P. Ha, Phys. Rev.D{\bf89}, no. 9,
094027 (2014).\\

\end{document}